\documentclass[twocolumn,amsmath,amssymb,nofootinbib,prl,superscriptaddress]{revtex4-1}
\pdfoutput=1

\usepackage{graphicx}
\usepackage{dcolumn}
\usepackage{bm,psfrag}
\usepackage{pdfpages}

\def\Bizon{Bizo\'n\, }

\newcommand{\be}{\begin{equation}}
\newcommand{\ee}{\end{equation}}
\newcommand{\bea}{\begin{eqnarray}}
\newcommand{\eea}{\end{eqnarray}}
\newcommand{\ba}{\begin{eqnarray}}
\newcommand{\ea}{\end{eqnarray}}

\newcommand{\beq}{\begin{equation}}
\newcommand{\eeq}{\end{equation}}
\newcommand{\beqa}{\begin{eqnarray}}
\newcommand{\eeqa}{\end{eqnarray}}
\newcommand{\beqar}{\begin{eqnarray*}}
\newcommand{\eeqar}{\end{eqnarray*}}

\newcommand{\eqlabel}[1]{\label{#1}}  

\def\t6 {T_\mt{D6}}

\newcommand{\mt}[1]{\textrm{\tiny #1}}

\def\cale         {{\cal E}}

\def\cals         {{\cal S}}

\def\del          {\partial}

\def\ee           {{\rm e}}

\def\sqr#1#2{{\vcenter{\vbox{\hrule height.#2pt
 \hbox{\vrule width.#2pt height#1pt \kern#1pt
 \vrule width.#2pt}\hrule height.#2pt}}}}

\def\a{\alpha}

\def\dd{\delta}

\def\ee{\cale}

\def\aa1{\phi}
\def\cc1{\psi}

\usepackage[bookmarks=false]{hyperref} 
\hypersetup{pdfstartview=FitH,pdfhighlight=/O,colorlinks=false}

\begin{document}

\preprint{arXiv:1402.nnnn [hep-th]; UWO-TH-14/xx}

\title{Holographic Thermalization, stability of AdS, and the Fermi-Pasta-Ulam-Tsingou paradox}


\author{Venkat Balasubramanian}
\email{vbalasu8@uwo.ca}
\affiliation{Department of Applied Mathematics, University of Western
Ontario, London, Ontario N6A 5B7, Canada}
\author{Alex Buchel}
\email{abuchel@perimeterinstitute.ca}
\affiliation{Department of Applied Mathematics, University of Western
Ontario, London, Ontario N6A 5B7, Canada}
\affiliation{Perimeter Institute for Theoretical Physics, Waterloo, Ontario N2L 2Y5,
Canada}
\author{Stephen R. Green}
\email{sgreen04@uoguelph.ca}
\thanks{CITA National Fellow}
\affiliation{Department of Physics, University of Guelph, Guelph, Ontario N1G 2W1, Canada}
\author{Luis Lehner}
\email{llehner@perimeterinstitute.ca}
\affiliation{Perimeter Institute for Theoretical Physics, Waterloo, Ontario N2L 2Y5,
Canada}
\author{Steven L. Liebling}
\email{steve.liebling@liu.edu}
\affiliation{Department of Physics, Long Island University, Brookville, NY 11548, U.S.A}

\begin{abstract}
  For a real massless scalar field in general relativity with a
  negative cosmological constant, we uncover a large class of
  spherically symmetric initial conditions that are close to AdS, but
  whose numerical evolution does not result in black hole formation.
  According to the AdS/CFT dictionary, these bulk solutions are dual
  to states of a strongly interacting boundary CFT that fail to
  thermalize at late times. Furthermore, as these states are not
  stationary, they define dynamical CFT configurations that do not
  equilibrate.  We develop a two-timescale perturbative formalism that
  captures both direct and inverse cascades of energy and agrees with
  our fully nonlinear evolutions in the appropriate regime.  We also
  show that this formalism admits a large class of quasi-periodic
  solutions.  Finally, we demonstrate a striking parallel between the
  dynamics of AdS and the classic Fermi-Pasta-Ulam-Tsingou problem.
\end{abstract}

\maketitle

\noindent {\it Introduction.---}The gauge theory-string theory
correspondence \cite{Aharony:1999ti} has become a valuable tool to
study nonequilibrium phenomena in strongly interacting QFTs
\cite{Chesler:2008hg,Balasubramanian:2010ce,Buchel:2013gba}.  In a
particular limit, this correspondence links general relativity in
$d+1$-dimensional asymptotically anti-de Sitter (AdS$_{d+1}$)
spacetimes with $d$-dimensional conformal field theories.  A
question of particular importance in field theory is to understand the
process of equilibration and thermalization.  This corresponds, in the
bulk, to collapse of an initial perturbation to a black hole.

In the first detailed analysis~\cite{Bizon:2011gg} of dynamics of
perturbations of global AdS$_4$, \Bizon and Rostworowski argued that
(except for special {\em nonresonant} initial data) the evolution of a
real, massless, spherically symmetric scalar field {\em always}
results in gravitational collapse, even for arbitrarily small initial
field amplitude $\epsilon$.  At the linear level, this system is
characterized by a normal mode spectrum with natural frequencies
$\omega_j=2j+3$.  Using weakly nonlinear perturbation theory, these
authors described the onset of instability as a result of resonant
interactions between the normal modes.  Because of the presence of a
vast number of resonances, they argued that this mechanism leads to a
{\em direct turbulent cascade} of energy to high mode numbers, making
gravitational collapse inevitable.  Higher mode numbers are more
sharply peaked, so this corresponds to an effect of gravitational
focusing.

The analysis of \cite{Bizon:2011gg} also showed that, for initial data
consisting of a single mode, the dominant effect of resonant
self-interaction could be absorbed into a constant shift in the
frequency of the mode.  (This time-periodic solution was 
confirmed to persist at higher nonlinear order
\cite{Maliborski:2013jca}.)  However, for two-mode initial data,
additional resonances are present that cannot be absorbed into
frequency shifts.  The result is secular growth of higher modes.

The turbulent cascade described in~\cite{Bizon:2011gg} is a beautiful
mechanism for thermalization of strongly coupled QFTs with holographic
gravitational duals.  However, it was recently pointed out that this
cascade argument breaks down if {\em all} modes are initially
populated, and the mode amplitudes fall off sufficiently rapidly for
high mode numbers~\cite{Buchel:2012uh}.  In this case, all resonant
effects may once again be absorbed into frequency shifts and black
hole collapse is avoided.  Low-lying modes have broadly distributed
bulk profiles. Thus, one might expect that if the initial scalar
profile is broadly distributed, its evolution might not result in
gravitational collapse (see
also~\cite{Maliborski:2014rma,Dias:2012tq,Abajo-Arrastia:2014fma}).
This prediction was verified numerically~\cite{Buchel:2013uba}. The
physical mechanism responsible for collapse/non-collapse of small
amplitude initial data is a competition between two effects:
gravitational focusing and nonlinear dispersion of
the propagating scalar field. If the former dominates, gravitational
collapse ensues~\cite{Bizon:2011gg}.  If the latter does, the system
evolves without approaching any identifiable static or stationary
solution---the perturbed boundary CFT neither thermalizes nor
equilibrates at late times~\cite{Buchel:2013uba}.

The perturbation theory of~\cite{Bizon:2011gg} cannot make
predictions at late times.  (The growth of secular terms in the
expansion causes a breakdown at time $t \propto 1/\epsilon^2$.) It
also does not properly take into account energy transfer between
modes.  In this Letter, we undertake a thorough analysis of the
dynamics of AdS by making use of a new perturbative formalism for
analyzing the effect of resonances on the evolution of this system
{\em that is valid for long times}. We also perform fully nonlinear GR
simulations (see~\cite{Buchel:2012uh,Buchel:2013uba} for details of
our numerical implementation and validation).  In the process we uncover a close
relationship between the dynamics of AdS and the famous
Fermi-Pasta-Ulam-Tsingou~(FPUT) problem~\cite{Fermi:1955:SNP,2008PhT....61a..55D}.  Our
formalism is based on a two-timescale approach~\cite{opac-b1091299},
where we introduce a new ``slow time'' $\tau=\epsilon^2 t$.  The
timescale $\tau$ characterizes energy transfers between modes, whereas
the ``fast time'' $t$ characterizes the original normal
modes. Importantly, this formalism allows one to study the system for
long times and examine energy transfer between modes. In the following
we describe the {\em Two Time Framework} (TTF) and determine a large
class of quasi-periodic solutions that extends the single-mode
periodic solutions of \cite{Bizon:2011gg,Maliborski:2013jca}.  These
solutions have finely tuned energy spectra such that the net energy
flow into each mode vanishes, and they appear to be stable to small
perturbations within both TTF and full numerical simulations.  We then
study the behavior of two-mode initial data of~\cite{Bizon:2011gg}
under both approaches.  Finally, we use the TTF equations to draw an
interesting parallel between scalar collapse in AdS and the FPUT
problem of thermalization of nonlinearly coupled
oscillators~\cite{Fermi:1955:SNP}.

\noindent{\it Model.---}Following \cite{Bizon:2011gg}, we consider a
self-gravitating, real scalar field $\phi$ in asymptotically AdS$_4$
spacetime.  Imposing spherical symmetry, the metric takes the form
\begin{equation}
ds^2=\frac{1}{\cos^2 x}\left(-A e^{-2\delta} dt^2+A^{-1} dx^2+\sin^2 x\ d\Omega^2\right),
\eqlabel{metric}
\end{equation}
where we set the asymptotic AdS radius to one. Spherical symmetry
implies that $A$, $\delta$ and $\phi$ are functions of time $t \in
(-\infty, \infty)$ and the radial coordinate $x\in [0,\frac \pi 2)$.

In terms of the variables $\Pi\equiv e^\delta\dot\phi/A$ and
$\Phi\equiv \phi'$, the equation of motion for $\phi$ is
\begin{align}\label{eq:phi}
  \ddot\phi ={}& \left(\dot{A}e^{-\delta}-A\dot{\delta}e^{-\delta}\right)\Pi + A^2e^{-2\delta}\Phi'  \\
  &+ \left(\frac{2}{\sin x\cos x}A^2e^{-2\delta}  + AA' e^{-2\delta} - A^2e^{-2\delta}\delta'\right )\Phi ,\nonumber
\end{align}
while the Einstein equation reduces to the constraints,
\begin{align}
  \label{eq:A}A'&=\frac{1+2\sin^2x}{\sin x\cos x}(1-A)+\sin x\cos x A\left(|\Phi|^2+|\Pi|^2\right),\\
  \label{eq:delta}\delta'&=-\sin x\cos x\left(|\Phi|^2+|\Pi|^2\right).
\end{align}

\noindent{\it Two Time Framework.---}TTF consists of defining the slow
time $\tau=\epsilon^2 t$ and expanding the fields as
\begin{align}
  \label{eq:phiexpansion}\phi &= \epsilon\phi_{(1)}(t,\tau,x) + \epsilon^{3}\phi_{(3)}(t,\tau,x) +O(\epsilon^5),\\
  \label{eq:Aexpansion}A &= 1 + \epsilon^2A_{(2)}(t,\tau,x) + O(\epsilon^4),\\
  \label{eq:deltaexpansion}\delta &= \epsilon^2\delta_{(2)}(t,\tau,x)+O(\epsilon^4).
\end{align}
It is possible to go beyond $O(\epsilon^3)$ by introducing additional
slow time variables.  However, the order of approximation used here is
sufficient to capture the key aspects of weakly nonlinear AdS collapse
in the $\epsilon\to 0$ limit.

Perturbative equations are derived by substituting the expansions
\eqref{eq:phiexpansion}--\eqref{eq:deltaexpansion} into the equations
of motion \eqref{eq:phi}--\eqref{eq:delta}, and equating powers of
$\epsilon$.  It is important to note that, when taking time
derivatives of a function of both time variables we have
$\partial_t \to \partial_t  + \epsilon^2 \partial_{\tau}$.
At $O(\epsilon)$, we obtain the wave equation for $\phi_{(1)}$
linearized off exact AdS,
\begin{equation}\label{eq:phi1}
  \partial_t^2{\phi}_{(1)} = \phi_{(1)}'' + \frac{2}{\sin x\cos x}\phi_{(1)}' \equiv -L\phi_{(1)}.
\end{equation}
The operator $L$ has eigenvalues $\omega_j^2=(2j+3)^2$ ($j=0,1,2,\ldots$)
and eigenvectors $e_j(x)$ (``oscillons'')~\cite{Bizon:2011gg}.  Explicitly,
\begin{equation}
  e_j(x)=d_j\cos^3 x\ _2 F_1\left(-j,3+j;\frac 32; \sin^2 x\right),
  \eqlabel{adef}
\end{equation}
with $d_j=4 \sqrt{(j+1) (j+2)}/\sqrt{\pi }$.  The oscillons form
an orthonormal basis under the inner product
\begin{equation}
  (f,g)=\int_0^{\pi/2}f(x)g(x)\tan^2x\,\mathrm{d}x.
\end{equation}
 The general real solution to \eqref{eq:phi1} is
\begin{equation}\label{eq:phi1sol}
  \phi_{(1)}(t,\tau,x) = \sum_{j=0}^\infty \left(A_j(\tau)e^{-i\omega_jt}+\bar{A}_j(\tau)e^{i\omega_j t}\right)e_j(x),
\end{equation}
where $A_j(\tau)$ are arbitrary functions of $\tau$, to be
determined later.

At $O(\epsilon^2)$ the constraints \eqref{eq:A}--\eqref{eq:delta} have solutions
\begin{align}
  \label{eq:A2}A_{(2)}(x) &= -\frac{\cos^3x}{\sin x}\int_0^x\left(|\Phi_{(1)}(y)|^2+|\Pi_{(1)}(y)|^2\right)\tan^2y \,\mathrm{d}y,\\
  \label{eq:delta2}\delta_{(2)}(x) &= -\int_0^x\left(|\Phi_{(1)}(y)|^2+|\Pi_{(1)}(y)|^2\right)\sin y\cos y \,\mathrm{d}y.
\end{align}

Finally, at $O(\epsilon^3)$ we obtain the equation for $\phi_{(3)}$,
\begin{equation}
 \partial_t^2\phi_{(3)} + L\phi_{(3)}
+ 2\partial_t\partial_\tau\phi_{(1)} =S_{(3)}(t,\tau,x),
\eqlabel{phi3}
\end{equation} 
where the source term is 
\begin{align}
  S_{(3)}={}&\del_t(A_{(2)}-\delta_{(2)})\del_t\phi_{(1)}-2(A_{(2)}-\dd_{(2)}) L\phi_{(1)}\nonumber\\
  &+(A_{(2)}'-\dd_2')\phi_{(1)}'.
\eqlabel{s3}
\end{align}
The solutions \eqref{eq:A2}--\eqref{eq:delta2} for $A_{(2)}$ and
$\delta_{(2)}$ are  substituted directly into $S_{(3)}$.  In
general, the source term $S_{(3)}$ contains resonant terms ({\em
  i.e.,} terms proportional to $e^{\pm i\omega_j t}$).  As noted in
\cite{Bizon:2011gg}, for all triads $(j_1,j_2,j_3)$, resonances occur
at $\omega_j=\omega_{j_1}+\omega_{j_2}-\omega_{j_3}$.  In ordinary
perturbation theory these resonances lead to secular growths in
$\phi_{(3)}$.  However,~\cite{Bizon:2011gg} showed that in {\em some}
cases the growths may be absorbed into frequency shifts.  TTF provides
a natural way to handle these resonances by taking advantage of the
new term $2\partial_t\partial_\tau\phi_{(1)}$ in \eqref{phi3} and the
freedom in $A_j(\tau)$.

We now project \eqref{phi3} onto an individual oscillon mode $e_j$ and
substitute for $\phi_{(1)}$,
\begin{align}
 &\quad\left(e_j,\partial_t^2 {\phi}_{(3)}+\omega_j^2\phi_{(3)}\right) 
-2i\omega_j\left(\partial_\tau A_je^{-i\omega_j t}-\partial_\tau{\bar{A}}_je^{i\omega_jt}\right)\nonumber\\
&= \left(e_j,S_{(3)}\right).
\eqlabel{project}
\end{align}
By exploiting the presence of terms proportional to $e^{\pm i\omega_jt}$
on the left hand side of the equation, we may cancel off the resonant
terms on the right hand side.  Denoting by $f[\omega_j]$ the part of
$f$ proportional to $e^{i\omega_j t}$, we set
\begin{align}
  -2i\omega_j \partial_\tau A_j =  (e_j,S(t,\tau,x))[-\omega_i]
  = \sum_{klm} \cals^{(j)}_{klm} \bar{A}_kA_lA_m,
\eqlabel{ttfequation}
\end{align}
where $\cals^{(j)}_{klm}$ are real constants representing different
resonance channel contributions.  The right hand side is a
cubic polynomial in $A_j$ and $\bar{A}_j$.  Thus, we have obtained a
set of coupled first order ODEs in $\tau$ for $A_j$,
which we shall refer to as the {\em TTF equations}.  The equations are
to be solved given the initial conditions for $\phi$.  This procedure
fixes the arbitrariness in the solution \eqref{eq:phi1sol} for
$\phi_{(1)}$.  While we could also solve for $\phi_{(3)}$,
this would be of little interest since the lack of resonances
remaining in \eqref{phi3} implies that $\phi_{(3)}$ remains bounded.

Under evolution via the TTF equations, both the amplitude and phase of
the complex coefficients $A_j(\tau)$ can vary.  Thus, in contrast to
the perturbative analysis in~\cite{Bizon:2011gg}, the energy per mode $E_j=\omega_j^2
|A_j|^2$ can change with time in a very nontrivial manner.  However,
it can be checked that the total energy $E=\sum_j E_j$ is conserved.
TTF thus describes an energy-conserving dynamical system.
The TTF equations also possess a scaling symmetry
$A_j(\tau) \to \epsilon A_j(\tau/\epsilon^2)$.  This symmetry was
observed in Fig.~2b of \cite{Bizon:2011gg}, which indicates that the 
instability mechanism is captured by TTF.

In practice, it is necessary to truncate the TTF equations at finite
$j=j_{\text{max}}$.  We evaluated  $\cals^{(j)}_{klm}$
up to $j_{\text{max}}=47$.  In particular, under truncation to
$j_{\text{max}}=0$, the equations reduce to
\begin{equation}
i\pi \partial_\tau A_0= 153 A_0^2\bar{A}_0,
\end{equation}
with solution $A_0(\tau)=A_0(0) \exp\left(-i\frac{153}{\pi}|A_0(0)|^2
  \tau\right)$.  This reproduces precisely the single-mode frequency
shift result of~\cite{Bizon:2011gg}.

\begin{figure}[h]
\begin{center}
 \includegraphics[width=3.2in,clip]{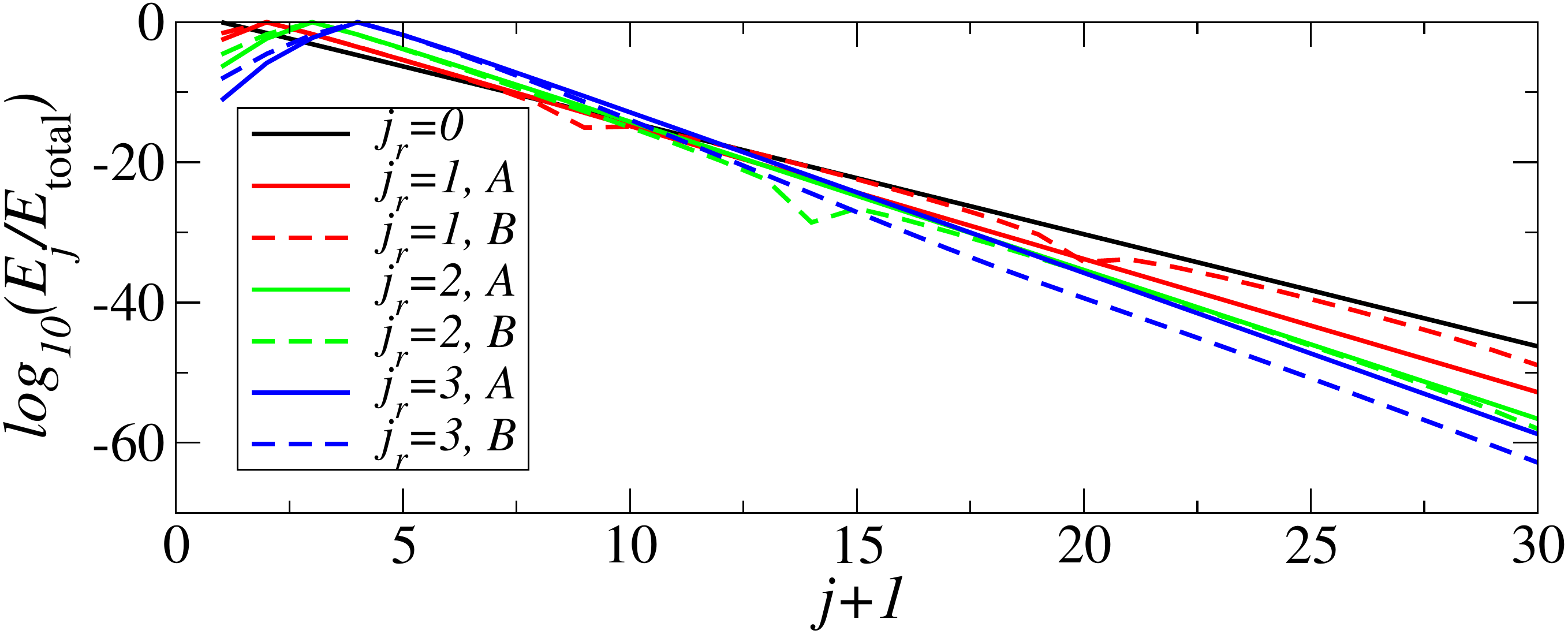}
\end{center}
\caption{Energy spectra of quasi-periodic solutions
  with ${\a_{j_{r}+1}}/{\a_{j_r}}=0.1$ for $j_r=0,1,2,3$. Dashed and
  solid lines distinguish different branches for $j_r>0$.  The solid
  branch is well-approximated by an exponential to each each side of
  $j_r$.  For ${\a_{j_{r}+1}}/{\a_{j_r}}$ too large, it becomes
  difficult to obtain solutions to~\eqref{eq:qp}, but for $j_r=0$, we
  can go up to ${\a_{1}}/{\a_{0}}\approx 0.42$.  (Constructed for
  $j_{\text{max}}=30$.)}
 \label{ttfpower}
\end{figure}

\noindent {\it Quasi-periodic solutions.---}To understand the dynamics
of TTF, we first look for quasi-periodic solutions.  For
$j_{\text{max}}=0$ this is the periodic solution above.  For general
$j_{\text{max}}>0$ we take as ansatz $A_j=\a_j \exp(-i\beta_j \tau)$,
where $\alpha_j,\beta_j\in\mathbb{R}$ are independent of $\tau$.
These solutions have $E_j=\text{constant}$, so they represent a
balancing of energy fluxes such that each mode has constant energy.
Substituting into the TTF equations, the $\tau$-dependence can be
canceled by requiring $\beta_j=\beta_0+j (\beta_1-\beta_0)$. This
leaves $j_{\text{max}}+1$ algebraic equations,
\begin{equation}\label{eq:qp}
  -2\omega_j \a_j \left[\beta_0+j (\beta_1-\beta_0)\right] = \sum_{kmn} \cals_{kmn}^{(j)} \a_k\a_m\a_n,
\end{equation}
for $j_{\text{max}}+3$ unknowns $(\beta_0,\beta_1,\{\alpha_j\})$.  The
equations for $j=0,1$ may be used to eliminate $(\beta_0,\beta_1)$,
leaving $j_{\text{max}}-1$ equations to be solved for
$\{\alpha_j\}$---two parameters of underdetermination.  The scaling
symmetry allows for elimination of one parameter, so we set
$\alpha_{j_r}=1$ for some fixed $0\le j_r <j_{\text{max}}$.  Taking
the remaining free parameter to be $\alpha_{j_r+1}$ and requiring
solutions to be insensitive to the value of $j_{\text{max}}$ ({\em
  i.e.}, stable to truncation), it is straightforward to construct
solutions perturbatively in $\a_{j_r+1}/\a_{j_r}$.  We find a single
solution for $j_r=0$ and precisely two otherwise (see
Fig.~\ref{ttfpower}).

\noindent{\it Stability of quasi-periodic
  solutions.---}Ref.~\cite{Maliborski:2013jca} extended single-mode,
time-periodic solutions to higher order in $\epsilon$ and found these
solutions to be stable to perturbations.  Similarly, we examine the
stability of our extended class of quasi-periodic solutions, both
using full numerical relativity simulations and by numerically solving
the TTF ODEs.

\begin{figure}[h]
\begin{center}
 \includegraphics[width=3.2in,clip]{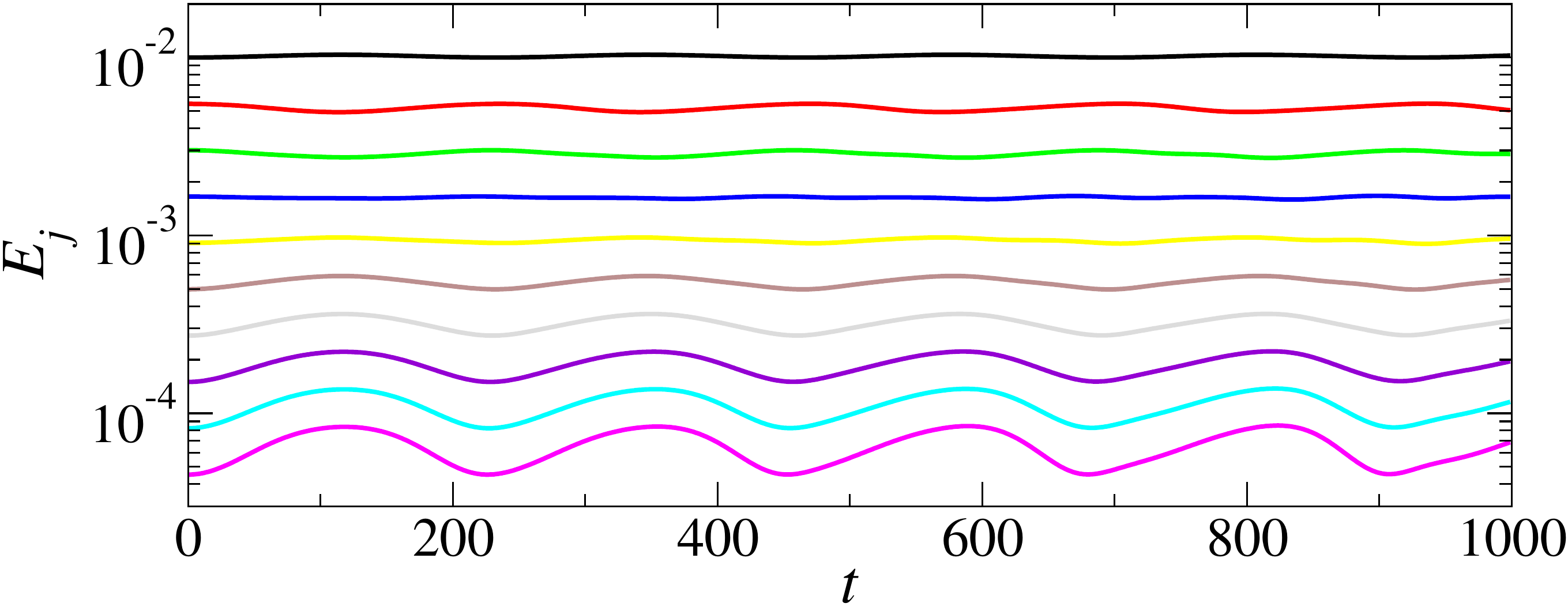}
\end{center}
\caption{Energy per mode for $0\le j\le 9$ for TTF solution with
  initial data $A_j(0) \propto \exp({-0.3 j})/(2j+3)$.}
 \label{fig:qp}
\end{figure}
We consider initial data $A_j(0) = \epsilon\exp({-\mu j})/(2j+3)$,
which well-approximates $j_r=0$ quasi-periodic solutions.  Varying
$\mu$ and also adding random perturbations, we observe periodic
oscillations about the quasi-periodic solution, providing evidence for
stability (see Fig.~\ref{fig:qp}).  For smaller values of $\mu$,
energy levels are more closely spaced, resulting in more rapid energy
transfers between modes, leading to larger-amplitude oscillations.
Likewise, larger random perturbations increase the amplitude of
oscillation, as the initial data deviates more strongly from a
quasi-periodic solution.  Results from TTF and full numerical
relativity simulations are in close agreement.

\noindent{\it Two-mode initial data.---}Our main interest is to
understand which initial conditions can be expected to collapse.  Thus
it is necessary to study initial data that are {\em not} expected to
closely approximate a quasi-periodic solution.  A particularly
interesting case consists of two modes initially excited (all others
zero) as this case was key to the argument of~\cite{Bizon:2011gg}
showing the onset of the turbulent cascade. In contrast to results of
the previous section, two-mode initial data,
\begin{equation}
A_j(0)=\frac{\epsilon}{3}\left(\delta_j^0+\kappa \delta_j^1\right),
\eqlabel{2mode}
\end{equation}
involves considerable energy transfer among modes provided $\kappa$ is
sufficiently large.  [For $\kappa\ll1$, \eqref{2mode} may be
considered as a perturbation about single-mode data.]  We examined
several choices of $\kappa$ using both TTF and full numerical
relativity, with similar results.  Here we restrict to
$\kappa=3/5$---the equal-energy case.

\begin{figure}[h]
\begin{center}
 \includegraphics[width=3.2in,clip]{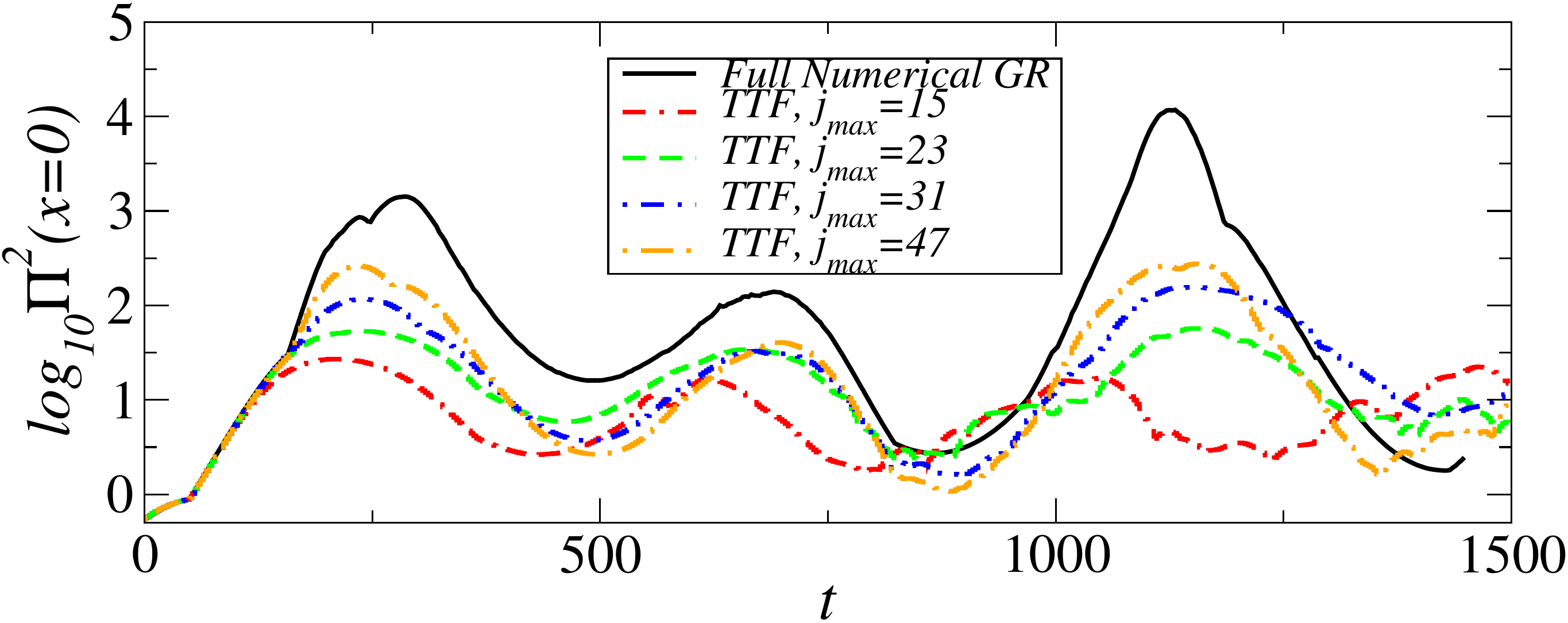}
\end{center}
\caption{Full numerical and TTF results for 2-mode equal-energy
  initial data with $\epsilon=0.09$. As $j_{\text{max}}$ is
  increased, the TTF solutions achieve better agreement with the full
  numerics.  Recurrence behavior observed in the full numerical
  solution is reasonably well captured by TTF.}
 \label{fig:2mode-convergence}
\end{figure}
The upper envelope of $\Pi^2(x=0)$ is often used as an indicator of
the onset of
instability~\cite{Bizon:2011gg,Buchel:2012uh,Buchel:2013uba}.  We plot
this quantity in Fig.~\ref{fig:2mode-convergence}, both for full GR
simulations and TTF solutions with varying $j_{\text{max}}$.  In the
full GR simulation, $\Pi^2(x=0)$ grows initially, but, in contrast to
blowup observed in~\cite{Bizon:2011gg} for Gaussian scalar field
profile, it then decreases close to its initial value. This {\em
  recurrence} phenomenon repeats and---for sufficiently small
$\epsilon$---collapse never occurs for as long as we have run the
simulation. Recurrence was also observed in previous work
\cite{Buchel:2013uba} for broadly distributed Gaussian profiles.

Also in Fig.~\ref{fig:2mode-convergence}, TTF solutions appear to
converge to the full numerical GR solution as $j_{\text{max}}$ is
increased.  (Strictly speaking, the TTF and numerical approaches converge as both
$j_{\text{max}} \rightarrow \infty$ and $\epsilon \rightarrow 0$; see the
accompanying supplemental material for more discussion.)
This illustrates nicely the cascade/collapse mechanism:
Higher-$j$ modes are more sharply peaked at $x=0$, so as the
(conserved) energy is transferred to these modes, $\Pi^2(x=0)$ attains
higher values.  Truncating the system at finite $j_{\text{max}}$
artificially places a bound on values of $\Pi^2(x=0)$ that can be
reached.  In particular, $\Pi^2(x=0)$ can never blow up for
$j_{\text{max}} < \infty$.


\begin{figure}[h]
\begin{center}
 \includegraphics[width=3in,clip]{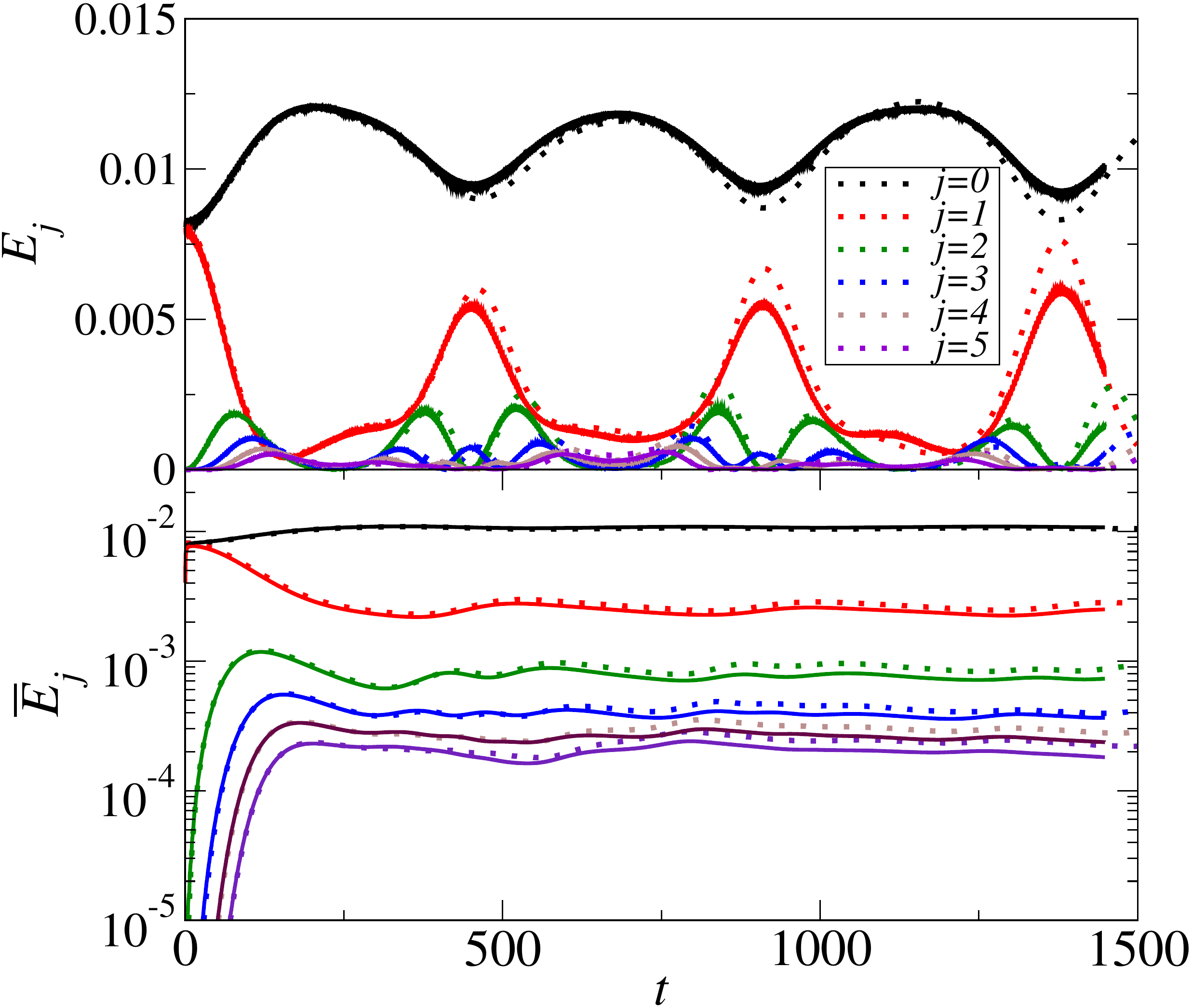}
\end{center}
\caption{Full numerical (solid) and TTF (dotted) energy (top panel)
  and running time-average energy (bottom panel) per mode, for 2-mode
  equal-energy initial data. Notice the repeated approximate return of
  the initial energies to the first two modes in the top panel and the
  running time-average energies asymptoting to distinct values.  For
  this run with $j_{\text{max}}=47$,
  $\sum_{j=0}^{11}|E_j^{\text{TTF}}-E_j^{\text{numerical}}|/E^{\text{total}}$
  does not exceed $0.19$.  For $j_{\text{max}}=(31,23,15)$ the bounds
  are $(0.28,0.42,0.57)$.  (The horizontal offset is partially attributed
  to a slight difference in time-normalization for our numerical and
  TTF codes.)}
 \label{fig:2mode-energy}
\end{figure}
It is useful to examine the solution mode by mode, and in
Fig.~\ref{fig:2mode-energy} we show the energy per mode as a function
of time.  Initially, energy is distributed evenly between modes
$j=0,1$.  It then flows out of $j=1$ to mode $j=2$, then $j=3$, etc.
At some point in time, energy begins to flow {\em back} to mode
$j=1$---an {\em inverse energy cascade}.  By $t\approx450$ the state
has nearly returned to the original configuration.  This recurrence
behavior then repeats.

The bottom plot of Fig.~\ref{fig:2mode-energy} illustrates the running time-average energy per mode
$\bar E_j(t) \equiv t^{-1} \int_0^{t} E_j(t') \,\mathrm{d}t'$.  Rather than
cascading to ever-higher modes, the energy sloshes primarily between
low-$j$ modes, in a ``metastable'' state.  We never observe
thermalization, {\em i.e.}, no equipartition of energy occurs.

Fig.~\ref{fig:2mode-energy} is remarkably similar in appearance to
plots of 
FPUT~\cite{Fermi:1955:SNP} (cf.~Figs.~4.1 and~4.2 of~\cite{fpubook}.)
FPUT numerically simulated a collection of nonlinearly coupled
harmonic oscillators and expected to see thermalization.  Instead,
they observed the same recurrence we see here.  Indeed, as the TTF
formulation~\eqref{ttfequation} of our system makes clear,
small-amplitude scalar collapse in AdS reduces precisely to a
(infinite) set of nonlinearly coupled oscillators, so the similar
behavior should not be surprising.  More precisely, our system is
related to the FPUT $\beta$-model \cite{fpubook}.  (Of course, the
particular resonances and nonlinear interactions differ between our
system and FPUT.)  Predicting when the FPUT system of oscillators
thermalizes is a longstanding problem in nonlinear dynamics, and is
indeed known as the FPUT {\em
  paradox}~\cite{fpubook,PhysRevE.52.307,2005Chaos..15a5104B}.

\noindent{\it Discussion.---}Common intuition suggests that a
finite-sized strongly interacting system driven off-equilibrium, even
by a small amount, eventually thermalizes. This thermalization would
imply, via AdS/CFT, that arbitrarily small perturbations about global
AdS {\em must} result in gravitational collapse.  However, we have
uncovered in this Letter a large class of initial conditions for a
massless, self-gravitating real scalar field in AdS$_4$, that {\em
  fail} to collapse.  We constructed and evolved these initial
conditions within a newly proposed TTF, as well as through full
numerical GR simulations.  TTF shows that scalar perturbations of AdS
are in the same universality class as the famous FPUT
problem~\cite{Fermi:1955:SNP}. Thus, perturbed AdS spacetimes act as a
holographic bridge between non-equilibrium dynamics of CFTs and the
dynamics of nonlinearly coupled oscillators and the FPUT paradox. In
this Letter we focused on the dynamics of low-energy 2+1 dimensional
CFT excitations ``prepared'' with nonzero expectation values of
dimension three (marginal) operators.  Extensions to
higher-dimensional CFTs, as well as to states generated by
(ir)relevant operators are straightforward.

\noindent {\bf Acknowledgments:} We would like to thank A. Polkovnikov
and J. Santos for interesting discussions and correspondence.  We also
thank D. Minic and A. Zimmerman for pointing us to the FPUT
problem. Finally, we thank P. Bizon and A. Rostworowski for
discussions and direct comparison of numerical solutions.  This work
was supported by the NSF under grants PHY-0969827 \&
PHY-1308621~(LIU), NASA under grant NNX13AH01G, NSERC through a
Discovery Grant (to A.B. and L.L.) and CIFAR (to L.L.).
S.R.G. acknowledges support by a CITA National Fellowship.  L.L. and
S.L.L.  also acknowledge the Centre for Theoretical Cosmology at the
University of Cambridge for their hospitality and its participants for
helpful discussions at the ``New Frontiers in Dynamical Gravity"
meeting during which this work was completed.  Research at Perimeter
Institute is supported through Industry Canada and by the Province of
Ontario through the Ministry of Research \& Innovation.  Computations
were performed at Sharcnet.

\bibliography{references}



\section*{Supplementary material}

# Holographic Thermalization, stability of AdS, and the Fermi-Pasta-Ulam-Tsingou paradox:
## Supplementary Material

Venkat Balasubramanian,[1, *] Alex Buchel,[1, 2, †] Stephen R. Green,[3, ‡] Luis Lehner,[2, §] and Steven L. Liebling[4, ¶]

[1]*Department of Applied Mathematics, University of Western Ontario, London, Ontario N6A 5B7, Canada*
[2]*Perimeter Institute for Theoretical Physics, Waterloo, Ontario N2L 2Y5, Canada*
[3]*Department of Physics, University of Guelph, Guelph, Ontario N1G 2W1, Canada*
[4]*Department of Physics, Long Island University, Brookville, NY 11548, U.S.A*

*Summary.*—We provide additional description of the approximations involved in TTF, as well as expectations for convergence in both $\epsilon$ and $j_{\max}$ to full general relativity. Last, we present results of tests validating our full GR code.

*Approximations.*—As described in our Letter, TTF involves two approximations: small $\epsilon$ and truncation in mode number $j$. By working to $O(\epsilon^3)$ we neglect interactions of higher than cubic order in metric perturbations, but we keep the lowest order nontrivial interactions. This gives rise to an infinite set of coupled oscillons, which in practice must be truncated at a finite mode number $j = j_{\max}$. We were able to explicitly determine the equations up to $j_{\max} = 47$.

*Small $\epsilon$.*—The TTF equations possess a scaling symmetry, $A_j(\tau) \to \epsilon A_j(\tau/\epsilon^2)$. The full Einstein equation does not possess this symmetry, however, and it is evident that had we included higher order in $\epsilon$ interactions in TTF, they would break the scaling symmetry. The presence of the scaling symmetry in a set of solutions to the full Einstein equation therefore indicates that higher order in $\epsilon$ interactions are negligible for these solutions. In Fig. 1 we plot the upper envelope over rapid oscillations of $\Pi^2(x=0)$ for full numerical solutions obtained with several values of $\epsilon$ for 2-mode equal-energy initial data, as well as a plot of rescaled values. It is clear that these solutions *do* possess an approximate scaling symmetry for sufficiently small $\epsilon$ (see also Fig. 2 of [1]).

Initial data for Fig. 1 is the same (up to overall normalization) as that of Figs. 3 and 4 of the Letter. Because of the approximate scaling symmetry in Fig. 1, the discrepancy between TTF and full GR in Fig. 3 of the Letter is therefore primarily attributed to truncation in mode number $j$. In addition, as $\epsilon$ is decreased, comparison of Fig. 1 and Fig. 3 of the Letter reveals that the small degree of non-scaling of the full GR solution brings it into closer agreement with TTF.

*Mode number truncation.*—For 2-mode equal-energy initial data, under evolution the majority of the energy remains in the lowest-$j$ modes (see Fig. 4 of the Letter). Nevertheless, $\Pi^2(x=0)$ can become very peaked. This is because higher-$j$ modes are more sharply peaked about the origin. Even small amounts of energy in high-$j$ modes can lead to growth of $\Pi^2(x=0)$, despite the dynamics being dominated by low-$j$ modes. This property can lead to

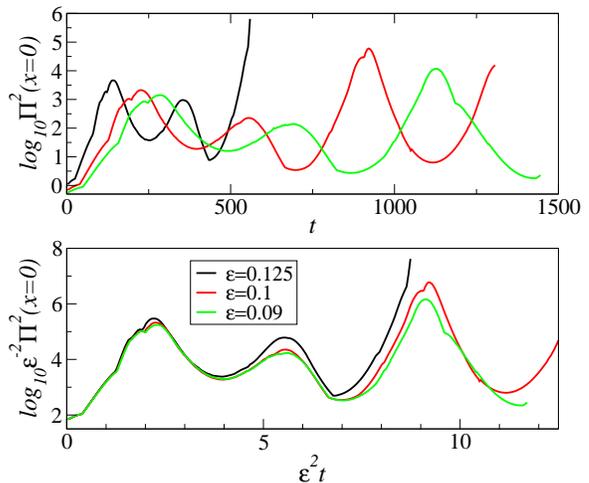

FIG. 1. (Color online) Full numerical GR solutions for 2-mode equal-energy initial data for different values of $\epsilon$. The bottom plot shows the same data, appropriately rescaled in time and amplitude by $\epsilon$. The $\epsilon = 0.125$ case collapses to black hole at $t \approx 550$. However, for smaller $\epsilon$ black hole formation is avoided with an inverse cascade. As $\epsilon$ increases and collapse to black hole ensues, the scaling symmetry in $\epsilon$ is broken as higher order terms (above third order) become important.

collapse in the case where sufficient energy is transferred to high-$j$ modes (*e.g.*, Gaussian initial data in [1]).

We have attempted to show in Fig. 3 of the Letter that increasing $j_{\max}$ gives rise to a better approximation of the full solution. This is best illustrated at early times, before the higher modes are populated, in which case the lack of available higher modes in a truncated TTF system is irrelevant. However, once the truncation begins to affect the dynamics, the overall solution is altered and it is more difficult to compare with full GR. Nevertheless, it is clear throughout the simulation that for higher $j_{\max}$, larger values of $\Pi^2(x=0)$ can be attained.

This discussion and the accompanying figures support the assertions that: (i) the TTF and the fully nonlinear solutions converge as $j_{\max} \to \infty$ and $\epsilon \to 0$, and (ii) the level to which the nonlinear solutions scale improves with decreasing $\epsilon$.

The truncated TTF equations are able to accurately model the dynamics of low-$j$ modes for the 2-mode data,

as shown in Fig. 4 of the main paper. However, the exact solution involves excitations of very high mode numbers at the peaks of $\Pi^2(x=0)$, and the TTF equations with $j_{\max} = 47$ are not sufficient to match the fully nonlinear result. However, we can instead consider a case where we expect better agreement. To that end, we include in Fig. 2 a plot of the full numerical GR and TTF solutions for an initially exponential energy spectrum. This solution is (a) close to a quasi-periodic solution, and (b) has very little energy in high-$j$ modes. In particular, the exponential fall-off in the energy spectrum with $j$ overwhelms the polynomial growth in peakiness about $x = 0$. Thus in this case, the truncated TTF and full numerical GR are in very close agreement[1].

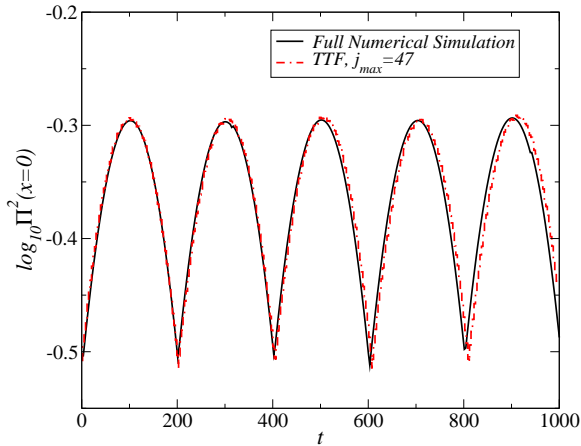

FIG. 2. (Color online) Full numerical GR and TTF solutions for initial data $A_j(0) = 0.05 \exp(-j)/(2j+3)$ which approximates a quasi-periodic solution.

*Full numerical GR code validation.*—The code we use has been described and thoroughly tested in [2, 3] to study scalar collapse in AdS. In particular we have confirmed: convergence of the obtained solutions with increased resolution, mass conservation, and convergence to zero of the constraint residuals (similar to Fig. 3 of [2]).

We present another, longer example of such a convergence study in Fig. 3. The results unambiguously indicate a convergent code over a large number of dynamical timescales (roughly $600/\pi \approx 190$). The test also directly confirms that, while a direct cascade dominates at early times, the system exhibits both direct and inverse cascades that compete with each other, a central result of this Letter; note in the bottom panel that the three highest resolutions have essentially converged to such a transition at $t \approx 230$.

This test also helps understand the failure of the TTF in Fig. 3 of the Letter to resolve the first peak of $\Pi^2(0, t)$. As mentioned previously, the TTF simply cannot resolve energy in high frequency modes compared to $j_{\max}$. Even the fully nonlinear code must extend to very high resolutions to resolve it spatially. Nevertheless, as indicated in Fig. 4 of the Letter and in Fig. 2 of this supplement, the TTF captures the essential dynamics of the system and matches the behavior of the fully nonlinear evolutions for the low frequency modes.

This study tests evolutions with a single grid-spacing, but in practice we use adaptive mesh refinement (AMR) to greatly reduce computational overhead. With an AMR code such as ours, one can vary a number of parameters to increase the effective resolution, and we often increase both the resolution of the base grid as well the criterion for refinement. We verify that such changes do not produce different results (see Figs. 5 and 6 of Ref. [3] for past examples of tests of our AMR code). In particular, AMR results are consistent with unigrid results that show a series of direct and indirect cascades (see Fig. 4).

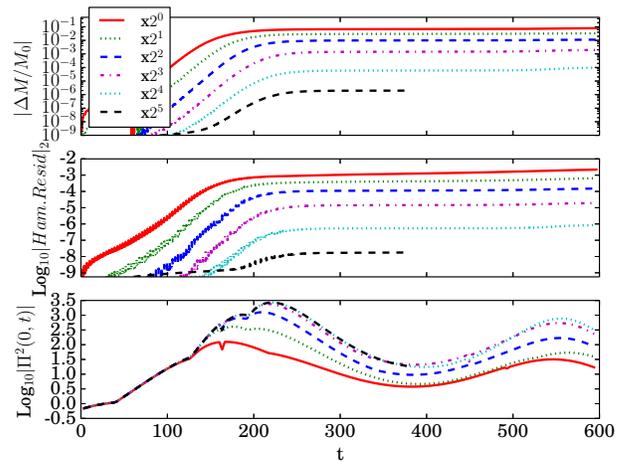

FIG. 3. (Color online) Demonstration of convergence for unigrid evolutions of equal-energy, two-mode initial data with $\epsilon = 0.1$. Shown for subsequent doublings of resolution are: (**top**) total mass loss, (**middle**) norm of the residual of the Hamiltonian constraint, and (**bottom**) the value of $\Pi^2(0, t)$. Note that the top panels display a measure of the error which progressively diminishes with resolution, indicating convergence. The bottom panel displays a feature of the evolution and quickly approaches a unique solution consistent with convergence. The base resolution (red, solid line) uses $316 + 1$ points while the highest resolution (black, dashed line) uses $10,112 + 1$ points.

As a final comment we note that in keeping with Ref. [4], we refer to the famous FPUT paradox within the

---

[1] The horizontal offset in Fig. 2 (and Figs. 3 and 4 of the Letter) between TTF and full numerical GR is attributed to a small difference in time-normalization between our codes. Specifically, the full numerical GR code sets $\delta = 0$ at the AdS boundary, whereas the TTF equations set $\delta = 0$ at $x = 0$. This gauge freedom amounts to a difference in time-normalization. The first nonzero contribution to $\delta$ is at $O(\epsilon^2)$, so this is a small effect.

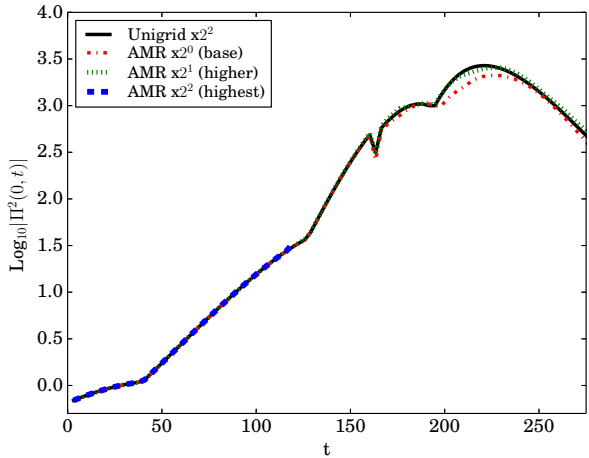

FIG. 4. (Color online) Demonstration of convergence for AMR evolutions of equal-energy, two-mode initial data with $\epsilon = 0.1$. Shown is the value of $\Pi^2(0,t)$ for three different AMR resolutions as well as the highest unigrid result from Fig. 3. Each successive AMR evolution uses a base grid with twice the resolution and a smaller threshold for refinement. The highest resolution AMR result has the same base resolution as the unigrid result.

main text by all four names of the contributors, Fermi, Pasta, Ulam, and Tsingou (later Menzel).

* vbalasu8@uwo.ca
† abuchel@perimeterinstitute.ca
‡ sgreen04@uoguelph.ca; CITA National Fellow
§ llehner@perimeterinstitute.ca
¶ steve.liebling@liu.edu



\end{document}